\documentclass[12pt]{iopart}   
\usepackage{graphicx}
\usepackage{epsfig}
\usepackage{amssymb}
\begin{document}

\newcommand{\be}{\begin{equation}} 
\newcommand{\ee}{\end{equation}}
\newcommand{\beqn}{\begin{eqnarray}} 
\newcommand{\eeqn}{\end{eqnarray}}

\title[Infinite randomness critical behavior of the contact process]{Infinite randomness critical behavior of the contact process on networks with long-range connections}

\author{R\'obert Juh\'asz}
\address{Institute for Solid
State Physics and Optics, Wigner Research Centre for Physics, H-1525 Budapest,
P.O. Box 49, Hungary}
\ead{juhasz.robert@wigner.mta.hu} 

\author{Istv\'an A. Kov\'acs}
\address{Institute for Solid
State Physics and Optics, Wigner Research Centre for Physics, H-1525 Budapest,
P.O. Box 49, Hungary}
\ead{kovacs.istvan@wigner.mta.hu} 

\begin{abstract}
The contact process and the slightly different susceptible-infected-susceptible model are studied on long-range connected networks 
in the presence of random transition rates by means
of a strong disorder renormalization group method and Monte Carlo
simulations. 
We focus on the case where the connection probability decays with the distance
$l$ as $p(l)\simeq\beta l^{-2}$ in one dimension. 
Here, the graph dimension of the network can be 
continuously tuned with $\beta$. 
The critical behavior of the models is found to be described 
by an infinite randomness fixed
point which manifests itself in logarithmic dynamical scaling. 
Estimates of the complete set of the critical exponents, which are found to vary with the
graph dimension, are provided by different methods. 
According to the results, the additional disorder of transition rates does not
alter the infinite randomness critical behavior induced by the disordered
topology of the underlying random network. 
This finding opens up the possibility of the application 
of an alternative tool, 
the strong disorder renormalization group method to dynamical processes 
on topologically disordered structures.  
As the random transverse-field Ising model falls into the same universality
class as the random contact process,  
the results can be directly transferred to that model 
defined on the same networks.

\end{abstract}

\maketitle

\section{Introduction}

Dynamical processes occurring on random networks have been in the 
center of interest recently \cite{BBV,dgm}. The motivation of this field is
many-sided, ranging from the spreading of epidemics or rumor in social
networks via spreading of computer viruses to transportation networks \cite{barabasi}. 
There are at least two striking properties of the underlying networks in these
problems which make the dynamics on the top of them much different from those
on regular lattices. 
First, the nodes are ``close'' to each other, meaning that the number of nodes
$n(\ell)$ that are reachable by traversing at most $\ell$ links from the
origin is increasing rapidly with $\ell$. In the case of an algebraic
relationship 
\be 
n(\ell)\sim \ell^{d_g},
\label{topdim}
\ee
a generalized {\it graph dimension} can 
be associated with the network under consideration. 
In many cases in the above systems the graph dimension of the underlying
network is formally infinite, like in the case of small world networks \cite{ws,newman} 
where, $\ln n(\ell)\sim \ell$, or in scale-free networks with a broad
distribution of degrees, where $\ln[\ln n(\ell)]\sim\ell$ \cite{barabasi}. 
This ``small-worldness'' makes mean-field approximations successful in the description of phase
transitions of spreading processes in many cases \cite{psv}.  
Second, the nodes of the underlying network are not equivalent since 
the degree of nodes, as well as the local neighborhoods are different and this 
circumstance may induce disorder effects on the dynamical process occurring on
it. 
This issue has been recently studied \cite{mjco,ojcm,jocm} in the case of a
paradigmatic model of an epidemic, the contact process \cite{cp}. 
In this model, there are binary variables attached to each node, which can be
either active or inactive (infected or healthy in the parlance of
epidemiology). Infected nodes then stochastically infect neighboring healthy
nodes or recover. Varying the relative rates of these two competing 
processes the system can
be driven from an active phase with a finite fraction of active nodes to an
absorbing one where all sites are inactive. The phase transition on a
translationally invariant lattice is known to fall into the directed
percolation universality class \cite{md,hhl,odor}. 
This model has been studied on
{\it long-range connected} (LRC) networks which consist of a $d$-dimensional 
hypercubic lattice and additional long-range edges existing
with probability that decays algebraically with their Euclidean length $l$ as 
$p(l)\simeq \beta l^{-s}$. 
This type of broad distribution of lengths of links
has been observed in a mobile phone communication network, 
where $s=2$ \cite{lambiotte} and in the
global airline network, where $s=3$ \cite{bpm}. 
The extremal cases $s=0$ and $s=\infty$
correspond to small-world networks and short-range networks,
respectively. 
In Refs. \cite{mjco,ojcm,jocm}, where the case $d=1$ was considered, 
the appearance of disorder effects 
was conjectured to be
related with the finiteness of the graph dimension of the underlying network. 
This is realized in the $d=1$ LRC network if the decay exponent $s$ is large enough, $s\ge 2$ namely, which
includes the ``critical'' point $s=2$ where the graph dimension varies
continuously with the prefactor $\beta$ of the asymptotical probability of
links \footnote{Note that, besides regular lattices, these networks provide an alternative for
approaching the limiting case $d_g=\infty$ through tuning the prefactor
$\beta$. Nevertheless, the two limiting cases may not be equivalent.}  \cite{bb,coppersmith,netrwre}. 
In this case, an anomalous slow, 
algebraic decay of the density of active 
nodes was observed in Monte Carlo simulations of the contact process in an extended phase on the
subcritical side of the transition point, at least for small enough graph
dimension.  
This region is analogous to the Griffiths-McCoy phase of 
disordered ferromagnets,
where the system, although it is globally paramagnetic, contains locally 
ferromagnetic domains of arbitrary size \cite{griffiths}.
In the corresponding phase of the contact process on LRC networks, 
the majority of the system is locally sub-critical 
but the randomness of the structure leads to
the formation of rare regions (sub-graphs) which contain an over-average
number of internal links and, as a consequence, may be locally
super-critical. The activity in these rare regions get extinct very slowly,
so they give a large contribution to the average density and result in
anomalous decay. 
The existence of this phase in the contact process with site-dependent random
rates on regular lattices has been known for a long time \cite{qcp}.
In the following, we will term this type of inhomogeneity as {\it parameter
  disorder} in order to distinguish it from {\it topological disorder} which
refers to the irregularities of the underlying network.  

Besides the off-critical behavior, parameter disorder has a striking effect on
critical scaling, as well \cite{moreira}, where, in the dynamical relations, 
the time is formally replaced by its logarithm.    
Indeed, an asymptotically exact {\it strong disorder renormalization group}
(SDRG) treatment \cite{im} of the one-dimensional model showed that, at least
for sufficiently strong disorder, the critical behavior is controlled by an 
{\it infinite randomness fixed-point} (IRFP) where the temporal scaling is
logarithmic and yielded the complete set of critical exponents \cite{hiv}. 
Later, this type of critical scaling has been indicated by the SDRG method in
higher dimensions $d=2,3,4$, as well as on the Erd\H os-R\'enyi graph \cite{er}
(where formally $d_g=\infty$) with parameter disorder \cite{ki,ki_alg,mg}.    
Based on these results, the existence of infinite-randomness critical behavior 
was conjectured on arbitrarily high-dimensional hypercubic lattices for sufficiently
strong parameter disorder.  

On LRC networks, where solely topological disorder is
present, numerical simulations have indicated a logarithmic critical 
scaling for small enough graph dimension, where the critical exponents 
vary with $d_g$ but a precise estimation of them is still lacking. 
Here, the critical point has been found to be flanked with a Griffiths phase, 
the width of which is shrinking with increasing $d_g$ \cite{mjco,ojcm,jocm}.  
For larger $d_g$ (from $d_g\approx2$ on) a Griffiths phase could not be observed and the
judgement of the critical behavior became less certain. Here the data were
compatible with a conventional algebraic scaling. 

In this paper, we wish to contribute to the above issue by revisiting some
still unclear parts, as well as to extend it to another direction.   
Namely, we shall focus on the critical behavior of the contact process on
$d=1$ LRC networks but, as a new feature, 
mainly {\it in the presence of parameter disorder}.
Keeping in mind the universality of the IRFP in case of 
parameter disorder, meaning that the critical
exponents are independent of the distribution of random parameters \cite{im}, 
we pose the
question whether, on a structure with topological disorder, the critical
exponents of the transition are modified by an additional parameter disorder
or not. 
An advantage of this model over that with homogeneous parameters is that it is suitable for a SDRG analysis, which always requires 
some initial parameter disorder. 
Furthermore, we expect the additional parameter disorder enhancing the
effective strength of disorder (i.e. reducing the transient regimes which are
usually rather long in the IRFP) and making possible a more accurate numerical
investigation.
We shall thus perform a numerical SDRG analysis of the above model and compare
the predictions with results of Monte Carlo simulations. 

It is worth mentioning that, provided it is controlled by an IRFP, the transition of the disordered
contact process falls into the same universality class as that of the random
transverse-field Ising model on any structure, as the strong disorder 
renormalization rules of the two models are identical (apart from factors that are irrelevant at criticality) \cite{fisher}. 
Besides regular lattices, the (zero-temperature) quantum phase transition in
this model has been studied also on a LRC network when the
underlying network is driven through a (long-range) percolation transition
\cite{dl}. Here, the critical exponents of the model can be expressed
in terms of those of the percolation transition. 
However, the nature of the quantum phase transition controlled by the strength
of the transverse field (on the percolating LRC network) rather than the percolation probability has not been revealed yet. 
This (much harder) problem corresponds to that formulated in the present work in the language of the contact process.
Due to universality, the results of our investigations can immediately be
transferred to the quantum critical behavior of the random transverse-field
Ising model.

The rest of the paper is organized as follows. 
The precise definition of the model will be given in section \ref{model}. 
In section \ref{scaling}, the scaling theory of the IRFP will be
recapitulated. Section \ref{sdrg} is devoted to the SDRG analysis of the
critical behavior, while results of Monte Carlo simulations are presented in
section \ref{montecarlo}. Finally, the results are discussed in section
\ref{discussion} and conclusions are drawn in section \ref{conclusions}.

\section{The model}
\label{model}

Networks with an algebraically decaying probability of long edges have been studied in the past from different aspects. 
In addition to the examples mentioned so far, they also arise as models of linear polymers with 
crosslinks between remote monomers \cite{chakrabarti}, in the context of 
decentralized search algorithms \cite{kleinberg}, or, indirectly, 
in susceptible-infected-recovered models with long-range infection \cite{grassberger}.  
Besides contact process \cite{mjco,ojcm,jocm}, 
long-range percolation \cite{an}, 
random walks \cite{rw,netrwre}, 
susceptible-infected-recovered model \cite{sir} 
and spin glass models \cite{sg} have been studied on them.
The geometry of these networks itself (the dependence of the diameter on the
number of nodes) has attracted much interest, as well \cite{bb,coppersmith,netrwre,sc,mam,grassberger,biskup,havlin}.

\begin{figure}[h]
\includegraphics[width=0.4\linewidth]{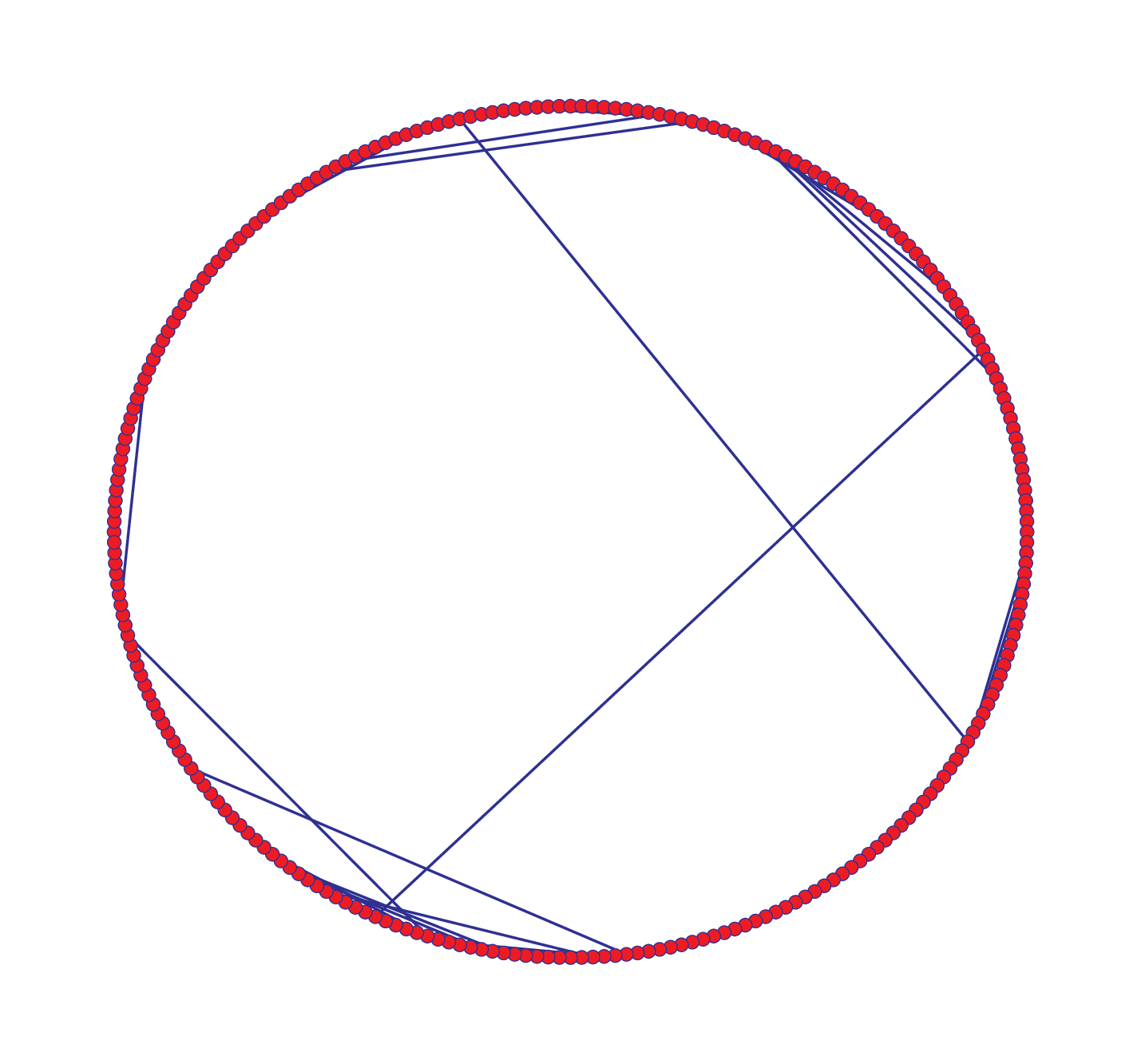}
\includegraphics[width=0.5\linewidth]{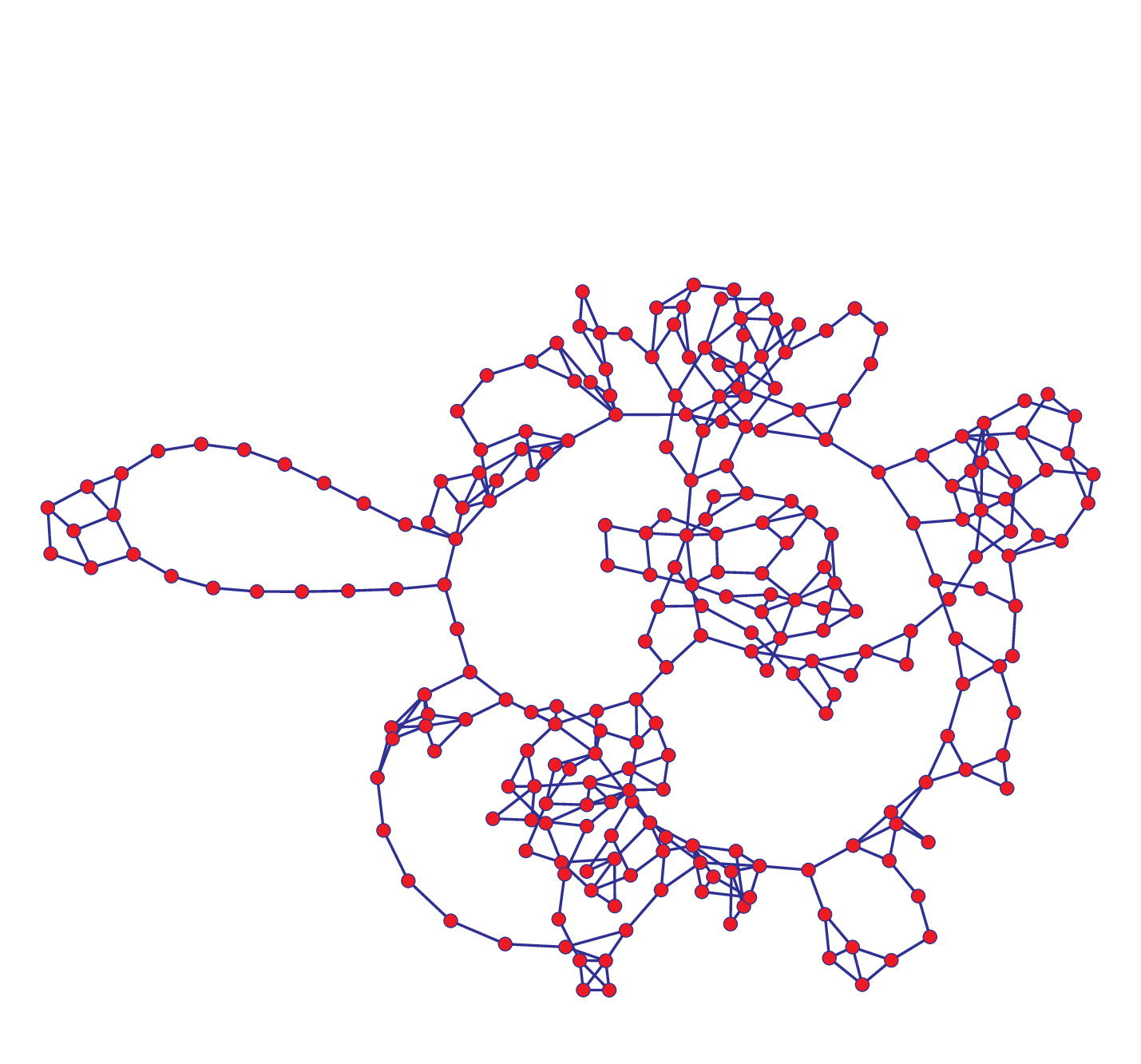}
\caption{
Left: A realization of the LRC networks with $N=256$ and $\beta=1$. 
Right: The same network represented with roughly equal edge lengths visualized by the Pajek program \cite{pajek}. 
\label{fig1} 
}
\end{figure}

Concerning the precise definition of the LRC networks, we shall adopt that of Ref. \cite{bb}. 
Let us have a set of $N$ nodes, which are labeled by integers $1,2,\dots,N$ and define the distance between node $i$ and $j$ as 
$l_{ij}=\min (|i-j|,N-|i-j|)$. 
This simply means that the nodes are arranged on a ring with unit 
spacing between them. 
Then all pairs of nodes with a distance $l_{ij}=1$ 
(i.e. neighboring nodes on the
ring) are connected with a link and 
all pairs with $l_{ij}>1$ are connected independently with the
probability
\be 
p(l)=1-\exp (-\beta l^{-s}), 
\label{pl}
\ee
where $\beta$ and $s$ are positive constants. 
For large $l$, this probability has the asymptotic form 
\be
p(l)\simeq \beta l^{-s}.
\label{pl_asymp}
\ee
A finite realization of this network is illustrated in Fig. \ref{fig1}. 

As for the geometry of these networks, 
for our purposes it is sufficient to know
that if $s=2$, the graph dimension is finite and  
continously increasing with $\beta$ from one
to infinity \cite{bb,coppersmith}, for numerical estimates, see
Refs. \cite{netrwre,grassberger}. 
If $s>2$, the network is short-ranged (i.e. $d_g=1$),
whereas if $s<2$, formally $d_g=\infty$ that hides in the range $1<s<2$ an
increase of the diameter with a power of $\ln N$ \cite{biskup}.  
In the present work, we shall restrict ourselves to the critical value $s=2$. 

To the nodes of these networks  binary variables $n_i$ coding active ($n_i=1$)
and inactive ($n_i=0$) states are attached, 
and on this state space a continuous-time 
stochastic process is considered with the following transitions. 
If node $i$ is active it can become inactive with a rate $\mu_i$, as well as it
tries to activate each neighboring node $j$ with a rate $\lambda_{ij}$.   
The rates $\mu_i$ and $\lambda_{ij}$ are i.i.d. quenched random 
variables
drawn from a distribution that will be specified later. 

Note that the homogeneous version of this model, where $\mu_i=\mu=\mathrm{const}$, 
$\lambda_{ij}=\mathrm{const}=\lambda$ is known as the susceptible-infected-susceptible
(SIS) model rather than the contact process. 
In the SIS model, the infection rate on links is constant whereas, in the contact process, the total rate of infection from nodes is constant.
In other words, in the
homogeneous contact process, the rate of infection from node $i$ through a
given link is $\lambda/d_i$, where $d_i$ is the coordination number 
(or degree) of node $i$. Therefore, in the case of the contact process on a non-regular lattice the
infection rates on a given link can be different in the two directions.
Although this situation is tractable by the SDRG method \cite{asym}, 
an efficient algorithm is at our disposal that works 
for the symmetric case, thus the SDRG analysis will be carried out for the SIS
model. 
In computer simulations, however, it is more natural to implement the
contact process rather than the SIS. 
(For the discrete time-implementation see section \ref{montecarlo}.) 
Nevertheless, in the case of networks with a narrow distribution of degrees,
which is the case for the LRC networks with an asymptotically $N$-independent 
distribution, the above distinction 
between the two models is irrelevant as far as the critical exponents are concerned. This will be demonstrated by results of numerical simulations in section \ref{montecarlo}.

\section{Scaling theory in an IRFP}
\label{scaling}

Here, following mainly Ref. \cite{hiv}, 
we shall briefly recapitulate the essentials of the scaling at an 
infinite randomness fixed-point, in a form adapted to finite-dimensional 
networks. 

As it is natural in strongly
disordered systems \cite{im}, we define the control parameter in the form 
$\Delta_0\equiv \overline{\ln(\lambda/\mu)}$, where the over-bar denotes
an average over the distribution of rates, whereas 
$\Delta\equiv \Delta_0-\Delta_0^*$ will denote the  
deviation from the critical value $\Delta_0^*$.   
As we will formulate the scaling theory for random networks, 
the usual linear size appearing in scaling relations will be 
replaced by the mean diameter $D$ of networks with $N$ nodes.  
The diameter of a network can be defined as the length $\ell$ of the longest shortest-path between any two nodes of the network, and, as can be gathered from Eq. (\ref{topdim}),  
its mean grows with $N$ as 
\be 
D(N)\sim N^{1/d_g}.
\label{DN}
\ee 

An appropriate order parameter of the transition is the probability $\rho$ 
that a randomly chosen node in the stationary state of the infinite system
($N\to\infty$) is active.
In the active phase and close to the critical point it vanishes as 
\be
\rho(\Delta)\sim\Delta^{\beta}, 
\ee
defining the order parameter exponent $\beta$. 
Close to the transition, the spatial correlation length (measured in terms of
the Euclidean distance $l$),
$\xi_{\perp}$ diverges as 
\be 
\xi_{\perp}\sim |\Delta|^{-\nu_{\perp}}, \qquad 
\ee
with the correlation length exponent $\nu_{\perp}$. 
The dynamics in an IRFP are strongly anisotropic; the time and length scales
are, namely, related to each other as 
\be
\ln\xi_{\parallel}\sim\xi_{\perp}^{\psi},
\label{act} 
\ee
where $\psi$ is the tunneling exponent and plays the role of a dynamical
exponent. Note that the conventional dynamical exponent that is defined by
an algebraic relationship rather than Eq. (\ref{act}) is formally infinite here.  
Based on the above, the order parameter for finite size $D$ and time $t$ is expected to have
the scaling form when the length (measured in shortest-path distance $\ell$) is rescaled by a factor $b$ as: 
\be
\rho(D,t,\Delta)=b^{-x}\tilde\rho(D/b,\ln t/b^{\psi},\Delta b^{1/\nu_{\perp}}),
\label{order}
\ee 
where $x\equiv \beta/\nu_{\perp}$.

The observables the time-dependences of which are usually measured in numerical
simulations starting from a single active node are
the survival probability, the average number of active nodes and the spread \cite{md}.  
The first one is the probability that there will be at least one active node at
time $t$. This probability is then averaged over the position of the initial
active seed. 
For the contact process, it scales in the same way as
the order parameter \cite{hhl,odor} so we have 
\be 
P(D,t,\Delta)=b^{-x}\tilde P(D/b,\ln t/b^{\psi},\Delta b^{1/\nu_{\perp}}). 
\label{P_scale} 
\ee  
Writing a scaling relation for the spatio-temporal correlation function
$C[n_0(t=0),n_i(t)]$
analogous to Eq. (\ref{order}) then summing over the position $i$ yields 
for the scaling of the average number of active nodes: 
\be  
N_a(D,t,\Delta)=b^{d_g-2x}\tilde N(D/b,\ln t/b^{\psi},\Delta b^{1/\nu_{\perp}}).\label{N_scale} 
\ee 
The spread is usually defined as the root-mean-square of the
Euclidean distance of active nodes from the origin. 
This quantity would, however, diverge in our model for $s\le 2$ owing to the
long links, therefore we define it in terms of the shortest-path distance
$\ell$ as 
\be 
R(t)=\sqrt{\overline{\langle \sum_in_i(t)\ell^2_i(t)/\sum_in_i(t)\rangle}},
\ee
where $\langle\cdot\rangle$ denotes the expected value conditioned on the
survival up to time $t$ for a given random
environment (i.e. a given random network and given starting position) whereas
the overbar denotes an average over the latter.  
Using again the scaling form of the spatio-temporal correlation function one
obtains that the spread obeys the scaling relation
\be  
R(D,t,\Delta)=b\tilde R(D/b,\ln t/b^{\psi},\Delta b^{1/\nu_{\perp}}).
\label{R_scale} 
\ee 

In the critical point ($\Delta=0$) of the infinite system ($D=\infty$) it
follows from the above relations that the observables depend on time asymptotically as 
\beqn 
P(t)\sim [\ln (t/t_0)]^{-\overline{\delta}}, \label{Pt} \\
N_a(t)\sim [\ln (t/t_0)]^{\overline{\eta}},  \label{Nt} \\
R(t)\sim [\ln(t/t_0)]^{1/\psi},
\label{Rt}
\eeqn
where $t_0$ is a non-universal microscopic time-scale and the exponents are given in terms of the earlier ones as 
\be
\overline{\delta}\equiv x/\psi, \qquad  
\overline{\eta}\equiv (d_g-2x)/\psi\;.
\label{deltax}
\ee 

A further quantity that is directly accessible by the SDRG method is the
lowest gap $\epsilon$ of the rate matrix (or infinitesimal generator) 
of the process in finite systems (see the next section). 
This quantity varies from sample-to-sample and can be interpreted as the
inverse of the mean time $\tau$ needed to reach the absorbing state starting
from the fully active one in a given finite sample. 
The distribution of its logarithm in samples with fixed $N$, in the
critical point $\Delta=0$ obeys the finite-size-scaling relation 
\be
\label{epsilon} 
f_{\epsilon}(\ln\epsilon,D)
=D^{-\psi}\tilde f_{\epsilon}(D^{-\psi}\ln\epsilon). 
\ee
Similarly, the distribution of mean lifetime $\tau$ has the scaling form 
\be 
f_{\tau}(\ln\tau,D)=D^{-\psi}\tilde f_{\tau}(D^{-\psi}\ln\tau). 
\label{taudist}
\ee

Note that, in the above scaling relations, we could have equally well used the
``volume'' $N$ of networks instead of their diameter $D$. But in that case, as
it is easy to see from Eq. (\ref{DN}), the
scaling exponents involving the finite size would differ from the present ones by a
factor of $1/d_g$ and the graph dimension in Eqs. (\ref{N_scale},\ref{deltax}) 
should be replaced by $1$. 
This set of exponents defined in terms of the volume will be distinguished by a prime from the above ones: 
\be 
\psi'\equiv \psi/d_g, \qquad 
x'\equiv x/d_g, \qquad 
\nu'\equiv \nu d_g. 
\label{prime}
\ee  
The other exponents are not effected by the choice of the measure of the size.

\section{SDRG treatment}
\label{sdrg}

By the SDRG procedure, the quickly relaxing degrees of freedom -- that are
related to high-lying levels of the rate matrix that governs the
time-evolution -- are sequentially eliminated, while lower-lying levels which are
responsible for the long-range dynamics are kept \cite{hiv}. 

To be specific, the procedure consists of two kinds of reduction steps. 
First, if the infection rate $\lambda_{ij}$ on a link is much greater than the
recovery rates on nodes $i$ and $j$ then the two nodes are merged and treated
as a single giant node with the effective recovery rate
\be
\tilde\mu\simeq 2\mu_i\mu_j/\lambda_{ij}
\label{rule1}
\ee
and size $\tilde m=m_i+m_j$. 
Second, if the recovery rate $\mu_i$ on node $i$ is much greater 
than the infection rates on the links emanating from it then node $i$ is
eliminated while any pairs of nodes neighboring to it are connected by new
links with effective infection rates  
\be
\tilde\lambda_{jk}\simeq \lambda_{ji}\lambda_{ik}/\mu_i
\label{rule2}
\ee
on them. 

In the original formulation of the SDRG scheme the largest
transition rate is selected then either of the above reduction steps are
applied and this loop is iterated until the system is sufficiently small so
that the spectrum of the rate matrix can be directly calculated.  
Although the reduction steps are approximative, they become more and more
accurate as the renormalization proceeds since the distributions of 
logarithmic rates broaden without limits, and ultimately, the
method becomes asymptotically exact in the IRFP \cite{fisher,im}. 

This renormalization scheme is formally equivalent to that of 
the random transverse-field Ising model defined by the Hamiltonian 
\be 
H=-\sum_{\langle ij\rangle}J_{ij}\sigma^x_i\sigma^x_j-\sum_ih_i\sigma^z_i,
\ee
where $\sigma_i^{x,z}$ are Pauli operators on site $i$, $J_{ij}$ and $h_i$ are
random couplings and external fields, respectively, and the first sum goes
over neighboring nodes of the network \cite{fisher}. 
The asymptotic renormalization rules of the model take the forms given 
in Eqs. (\ref{rule1},\ref{rule2}) with the 
correspondences $\mu_i\leftrightarrow h_i$ 
and $\lambda_{ij}\leftrightarrow J_{ij}$, 
apart from the absence of the factor of $2$ in Eq. (\ref{rule1}), which,
however, does not influence the critical exponents in an IRFP. 

The performance of the SDRG method in the above form may be rather ineffective
in other than one dimension since the connectedness of the underlying 
network will rapidly increase by the renormalization. 
To avoid this we will use a more efficient algorithm developed by one of us
\cite{ki_alg,phd}, which is based on the so called 'maximum rule'. 
According to this,  if multiple links between nodes would appear during the
renormalization, only the one with the maximal rate is kept. 
Application of this rule is thought not to influence the critical exponents 
in an IRFP and results in simplifications of the SDRG procedure.
The improved algorithm works by merely deleting links (and changing the rates
on the remaining ones), without generating new ones at all. The results of the
algorithm are identical to that of any traditional implementation of the SDRG
method (having also the widely applied maximum rule) for any finite graphs,
with $N$ sites and $E$ edges. However, we gain considerable time in
performance: while the traditional method needs $\mathcal{O}(N^3)$ operations,
the improved algorithm requires only $\mathcal{O}(N \log N + E)$, which is in practice much faster, than the $\mathcal{O}(N^2)$ operations needed to generate the LRC networks. 
The essence of the algorithm is that if two (or more) sources are able to
infect each other mutually before being healthy again, than these form a new
effective infection source, for which it takes a longer time to become healthy
again. So, in this sense, the process can be regarded as a 
special kind of percolation of the infection sources with a positive feedback \cite{ki_alg,phd}.

In the numerical SDRG analysis, both variables $\lambda_{ij}$ and $\mu_i$ where
taken from uniform distributions with probability densities 
$f_{\lambda}(\lambda)=\Theta(\lambda)\Theta(1-\lambda)$ and 
$f_{\mu}(\mu)=\frac{1}{\mu_m} \Theta(\mu)\Theta(\mu_m-\mu)$, respectively, 
where $\Theta(x)$ is the Heaviside step-function and we used the logarithmic variable, $\Delta_0=\ln(\mu_m)$ as a control parameter. 

\begin{figure}[!ht]
\begin{center}
\includegraphics[width=3.6in,angle=0]{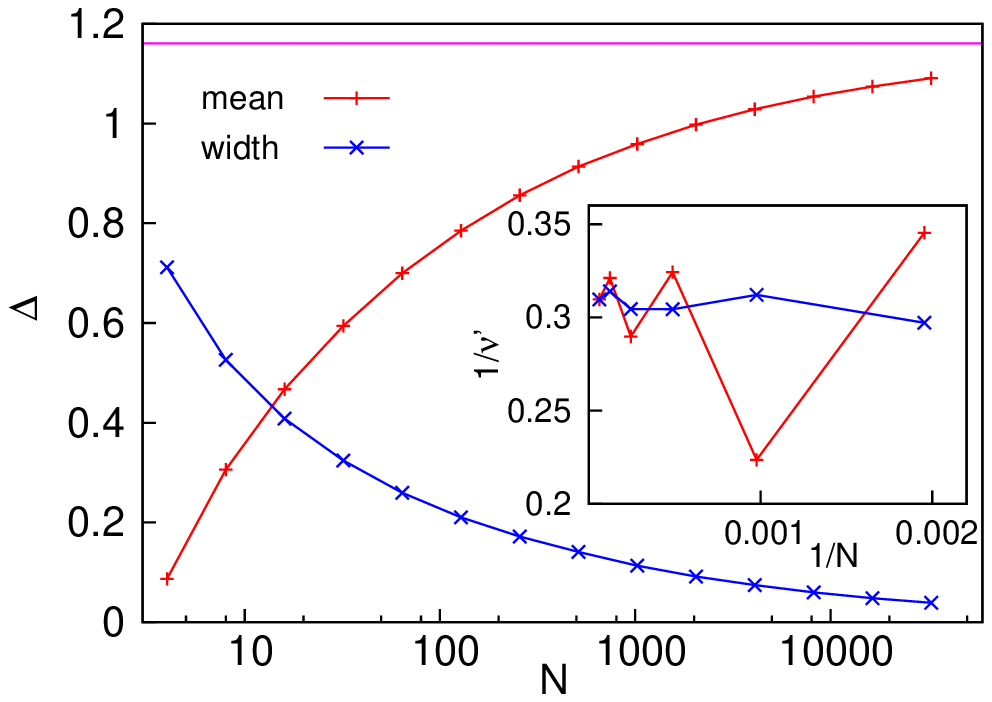}
\end{center}
\caption{
\label{fig_2} The distribution of the pseudo-critical points at
$\beta=0.75$ is illustrated by the mean value and the width of the
distribution for various values of the system size $N$. The extrapolated value
of the critical point is $\Delta^*_0=1.1606(15)$, indicated by the solid
line. Inset: The obtained finite-size estimates of the exponents $1/\nu'$ lead to the asymptotic value $1/\nu'=0.308(8)$.}
\end{figure}

It is generally a challenging task to precisely locate the critical point, and
the accuracy of the obtained critical exponents depends crucially on it. In
order to reach a sufficient precision, we have first determined the location
$\Delta_c$ of the pseudo-critical point for each random sample, where the correlation length reaches the size of the system. 
This can be conveniently obtained by the doubling method \cite{ladder,phd}. 
Here, the original network that has been built on a ring of nodes
is first made ``open'' by cutting all links
going over a given position $x$.   
Then two identical copies of this object are arranged in a ring of size $2L$ 
and the corresponding ``dangling'' links (that had been cut previously) are glued together. In this way, one obtains a network of size $2L$, which has a period $L$ along the ring. 
Now, $\Delta_c$ is given by the threshold value,
above which the last remaining giant nodes (or clusters) of the replicas fuse
together during the SDRG process. $\Delta_c$ varies from sample to sample, but
from the size-dependence of its distribution both the location $\Delta^*_0$ of
the 'true' critical point and the correlation length exponent $\nu'$ 
can be obtained according to the scaling form:
\be 
f_{\Delta}(\Delta,N)
=N^{1/\nu'}\tilde f_{\Delta}(\Delta N^{1/\nu'}), 
\ee
where $\Delta=\Delta_c-\Delta^*_0$. In practice, 
we study the scaling of the width of the distribution (given by the standard deviation), which is proportional to $N^{-1/d_g\nu}$ and the mean value, $|\Delta^*_0-\overline{\Delta_c}| \sim N^{-1/d_g\nu}$, and calculate size-dependent effective exponents by two-point and three-point fits, which are then extrapolated as $1/N\to0$, see Fig. \ref{fig_2}. 

Having at hand an accurate estimate for the location of the critical point,
the remaining two independent critical exponents ($x'$ and $\psi'$) can be
determined by applying the SDRG method at the critical point and analyzing 
the resulting cluster structure.
The gap $\epsilon$ of the rate matrix (or energy gap in the language of the
Ising model) is given by the effective recovery rate of the the last 
decimated giant node (cluster). The mass $m$ of the latter gives the magnetic moment of
the sample in  the corresponding Ising model, for details, 
see \cite{ki,ki_alg,phd}.
The number of independent random samples used in the calculations were at least $40000$, while the largest analyzed system size was typically $N=2^{15}$.

\begin{figure}[!ht]
\begin{center}
\includegraphics[width=3.6in,angle=0]{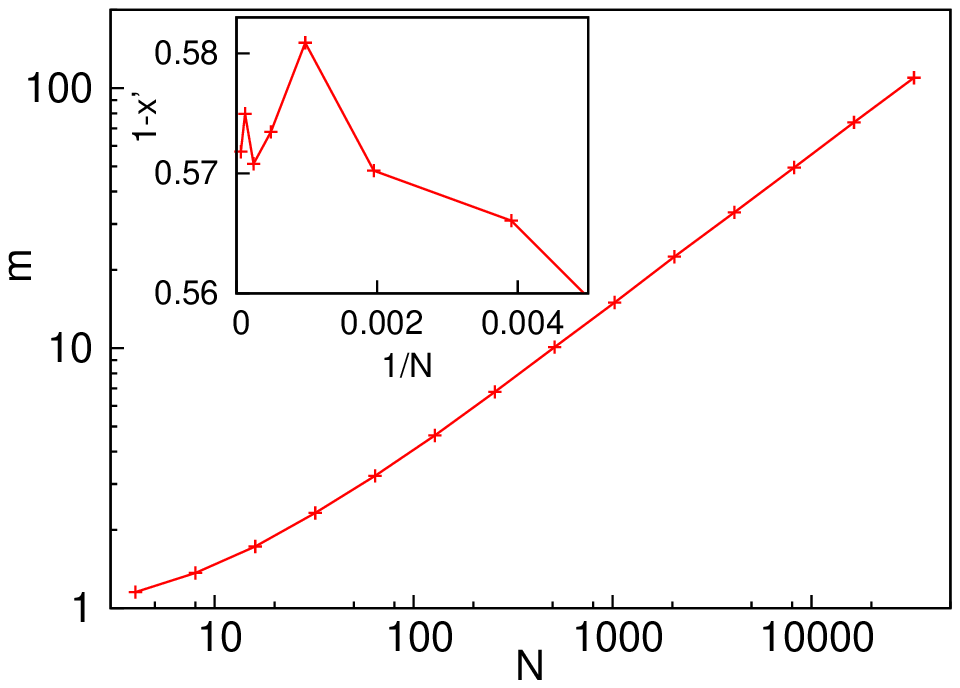}
\end{center}
\caption{
\label{fig_3} The average mass of the last cluster as the function of the system size, $N$ at $\beta=0.75$. Inset: The obtained finite-size estimates of $1-x'$ lead to the asymptotic value $1-x'=0.575(10)$.}
\end{figure}

According to the scaling theory presented in the previos section, 
the average mass scales with $N$ as $m(N) \sim N^{1-x'}$, 
since it is related to the order parameter 
(magnetization of the Ising model) $\rho$ as $\rho=m/N$. 
This is illustrated in  Fig.\ref{fig_3} for $\beta=0.75$. 
For sufficiently large system sizes, the points fit well to a straight
line. From two-point fits we can obtain finite-size estimates of $1-x'$,
which is then extrapolated as $1/N\to 0$.

\begin{figure}[!ht]
\begin{center}
\includegraphics[width=3.6in,angle=0]{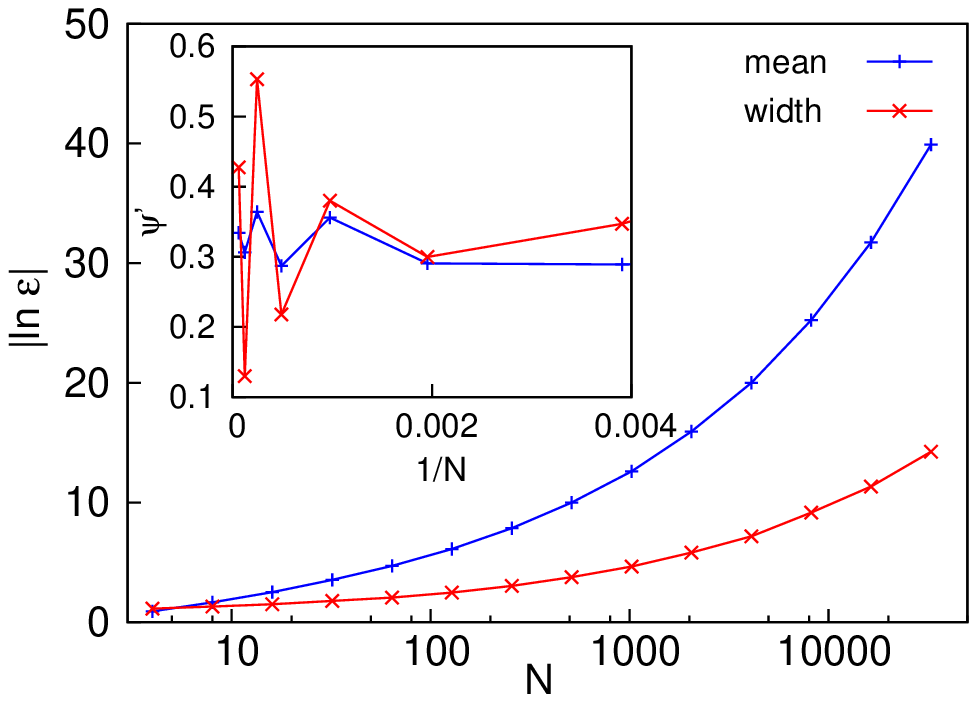}
\end{center}
\caption{
\label{fig_4} The mean value and width of the distribution of
the logarithmic gap plotted against the system size $N$ for $\beta=0.75$. 
Inset: The obtained finite-size estimates of $\psi'$ lead to the
asymptotic value $\psi'=0.32(3)$.}
\end{figure}

The mean value and the width of the distribution of $\ln\epsilon$ is shown
in Fig.\ref{fig_4} for $\beta=0.75$. As a clear indication of infinite
disorder scaling, the width of the distribution is increasing with
$N$. According to the scaling form in Eq. \ref{epsilon}, both the mean value
and the width is asymptotically proportional to $N^{\psi'}$. Similarly to
the other exponents, we determined finite-size estimates for $\psi'$ through
two-point fits, as illustrated in the inset of Fig. \ref{fig_4}. 
The estimates obtained by extrapolations to $N\to\infty$ are presented in Table \ref{tableRG}. 

\begin{table}[h]
\begin{center}
\begin{tabular}{|l||l|l|l|l|l|l|}
\hline   $\beta$&$d_g$&$\Delta^*_0$&$1/\nu'$ &$x'$&$\psi'$&$\psi$   \\
\hline
\hline   0.1  & 1.104(2) & 0.1738(4) & 0.422(7) & 0.214(3) &
0.524(30)& 0.58(3) \\
\hline   0.2  & 1.212(4) & 0.3505(5) & 0.366(4) & 0.231(5) & 0.53(3)
& 0.64(4) \\
\hline   0.3  & 1.353(7) & 0.529(2)  & 0.323(10)& 0.264(6)&
0.525(20)& 0.71(3) \\
\hline   0.4  & 1.499(7) & 0.700(1)  & 0.301(6) & 0.291(10)  & 0.485(30)
& 0.73(5) \\
\hline   0.5  & 1.656(8) & 0.856(1)  & 0.295(8) & 0.329(7)&
0.448(35)& 0.74(6) \\
\hline   0.75 & 2.03(2)  & 1.1606(15)& 0.308(8) & 0.426(8)& 0.33(3)
& 0.67(6) \\
\hline   1.0  & 2.347(17)& 1.3685(10)& 0.333(10)& 0.469(6) &
0.277(17)& 0.65(4) \\
\hline   2.0  & 3.045(27)& 1.853(3)  & 0.335(4) & 0.531(10)& 0.22(4)
& 0.67(12)\\
\hline
\end{tabular}
\end{center}
\caption{\label{tableRG} 
Critical exponents estimated by the SDRG method for different values of $\beta$. 
The graph dimensions are taken from Ref. \cite{netrwre}; for more precise
estimates, see Ref. \cite{grassberger}.}
\end{table}

\section{Monte Carlo simulations}
\label{montecarlo}

We have performed Monte Carlo simulations of the contact process on LRC
networks, implemented in the following way. 
A finite LRC network with $N$ nodes with  i.i.d. random variables 
$w_{ij}=w_{ji}$ 
on each link have been generated. 
Choosing an active node ($i$) randomly, it is set to
inactive with a probability  $1/(1+\lambda)$. 
Otherwise, a neighboring node ($j$) is chosen equiprobably and it is 
activated with a probability $w_{ij}$ (provided it was previously inactive).  
The random variables $w_{ij}$ have been drawn from a discrete, 
binary distribution with probability density 
\be
f(w)=c\delta(w-w_0)+(1-c)\delta(w-1).
\label{wdist}
\ee 
This can be interpreted in a way that a fraction $c$ of the links 
has a reduced capacity of transmitting the disease that is 
characterized by the parameter $w_0<1$. 
The parameters of this distribution
have been chosen to be $c=0.5$ and $w_0=0.2$ throughout the numerical 
simulations, except of the homogeneous model (i.e. the model without parameter
disorder) where formally $c=0$.
One Monte Carlo step of unit time consists of $N_a(t)$ such updates where 
$N_a(t)$ is the number of active nodes at the beginning of the step. We have generated random networks with a fixed $\beta$ in the range $[0.2,2]$
and with $N=1-5\cdot 10^6$ nodes (larger ones for larger $d_g$) and --
starting with a single active seed -- we have simulated
the process and measured the survival probability,
the number of active nodes and the spread as a function of time, 
typically, for MC times up to $2^{20}$.  
This measurement has been repeated for a fixed $\beta$ and $\lambda$ 
in $10^2$ independent random networks,
starting the process from $10^4$ different nodes per sample, and the measured
data have been averaged.   
When presenting results of the Monte Carlo simulations, we 
will simply use $\lambda$ as a control parameter rather than 
$\Delta_0$ defined in section \ref{scaling}.

\begin{figure}[h]
\includegraphics[width=0.5\linewidth]{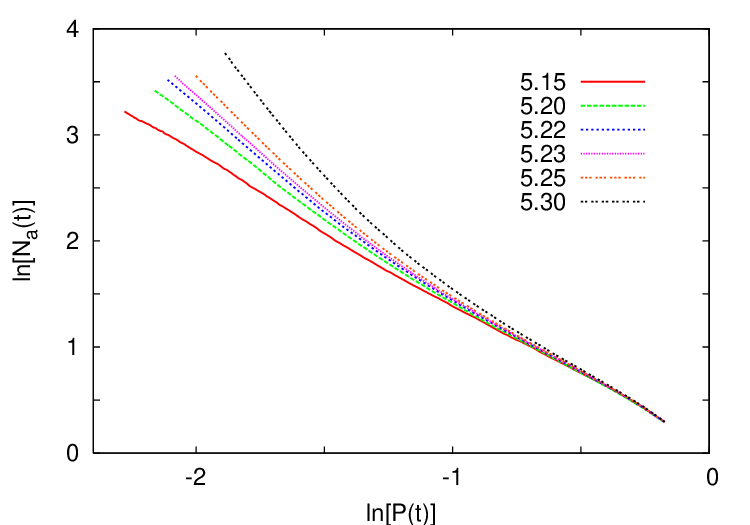}
\includegraphics[width=0.5\linewidth]{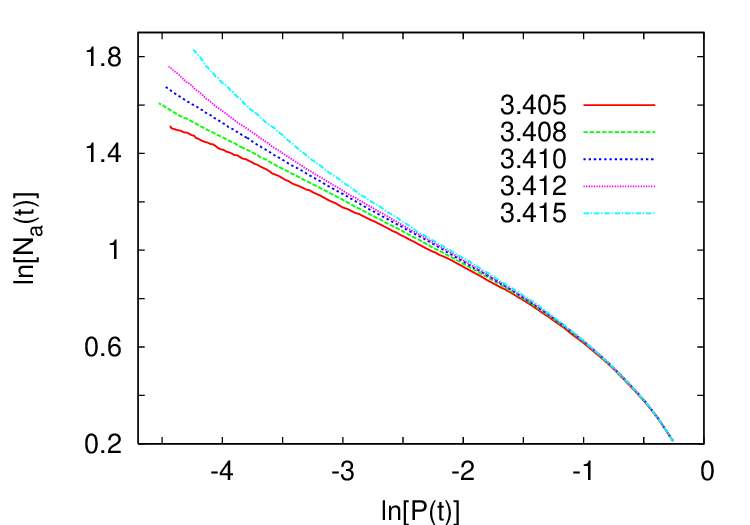}
\caption{
The logarithm of the average number of active nodes plotted against the
logarithm of the survival probability for different values of $\lambda$  for 
$\beta=0.3$ (left) and $\beta=1$ (right). 
The critical point is located at $\lambda^*=5.22(1)$ for
$\beta=0.3$ and at $\lambda^*=3.408(3)$ for $\beta=1$.  
\label{NP} 
}
\end{figure}

\begin{figure}[h]
\begin{center}
\includegraphics[width=0.5\linewidth]{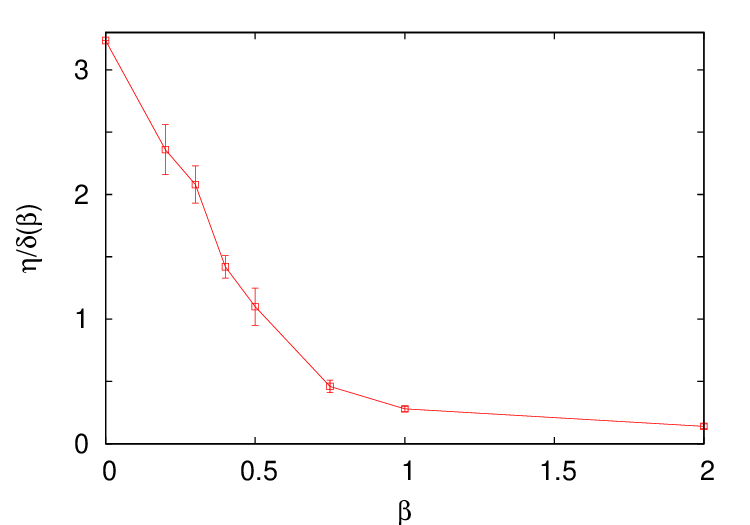}
\end{center}
\caption{
The ratio of exponents $\overline{\eta}/\overline{\delta}$ measured in Monte Carlo simulations for different values of $\beta$. 
The value at $\beta=0$ is the SDRG prediction for the one-dimensional random
contact process \cite{hiv}.  
\label{edx} 
}
\end{figure}

The critical point have been estimated in the way proposed in Ref. \cite{vfm} 
in order to avoid the difficulties about the large time scale $t_0$ in
Eqs. (\ref{Nt}).    
Plotting $\ln N_a(t)$ against $\ln P(t)$, the slope of the curve must tend to
$+1$ in the inactive phase, to $-\overline{\eta}/\overline{\delta}$ in the
critical point, and to $-\infty$ in the active phase, see Fig. \ref{NP}.   
Having identified the critical point for different $\beta$, 
we have determined $\overline{\eta}/\overline{\delta}$ by a linear fit to the
critical curve, see Fig. \ref{edx}. 
Using Eqs. (\ref{deltax},\ref{prime}), the exponent  $x'$ has then been 
calculated from 
\be 
x'=1/(2+\overline{\eta}/\overline{\delta}).
\ee
The estimates obtained this way are given in Table \ref{table}.  
Note that the most significant source of the error comes from the
uncertainty of the critical point. 

\begin{table}
\begin{center}
\begin{tabular}{|l||l|l|l|l|l|l|l|l|}
\hline $\beta$ &$d_g$ &$\lambda^*$&$\overline{\eta}/\overline{\delta}$& $x'$&$\psi$&
$\lambda^*$&$\overline{\eta}/\overline{\delta}$&$\psi$\\
& & & & & &($c=0$) &($c=0$) &($c=0$) \\
\hline
\hline   0.2  & 1.212(4) & 5.83(1) &2.36(20) &0.229(10)&0.57(5)&2.855(10)&2.3(3) &0.55(5)\\
\hline   0.3  & 1.353(7) & 5.22(1) &2.08(15) &0.245(9) &       &        &        &    \\
\hline   0.4  & 1.499(7) & 4.70(1) &1.42(9)  &0.292(8) &       &        &        &    \\
\hline   0.5  & 1.656(8) & 4.32(1) &1.10(15) &0.323(16)&0.56(8)&        &        &    \\
\hline   0.75 & 2.03(2)  & 3.728(5)&0.46(5)  &0.407(8) &0.60(8)&2.116(1)&0.45(5) &     \\
\hline   1.0  & 2.347(17)& 3.408(3)&0.28(2)  &0.439(4) &0.53(5)&1.976(1)&0.26(5) &0.52(7)\\
\hline   2.0  & 3.045(27)& 2.850(5)&0.14(2)  &0.467(4) &       &        &        &   \\
\hline
\end{tabular}
\end{center}
\caption{\label{table} 
Critical exponents estimated by Monte Carlo simulations
for different values of $\beta$. 
The graph dimensions are taken from Ref. \cite{netrwre}. Estimates in the
absence of parameter disorder ($c=0$) are shown in the last three columns.}
\end{table}

\begin{figure}[h]
\includegraphics[width=0.5\linewidth]{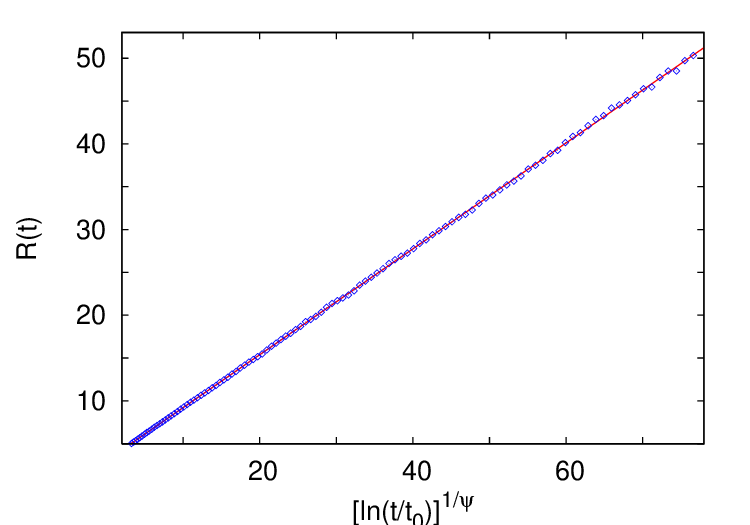}
\includegraphics[width=0.5\linewidth]{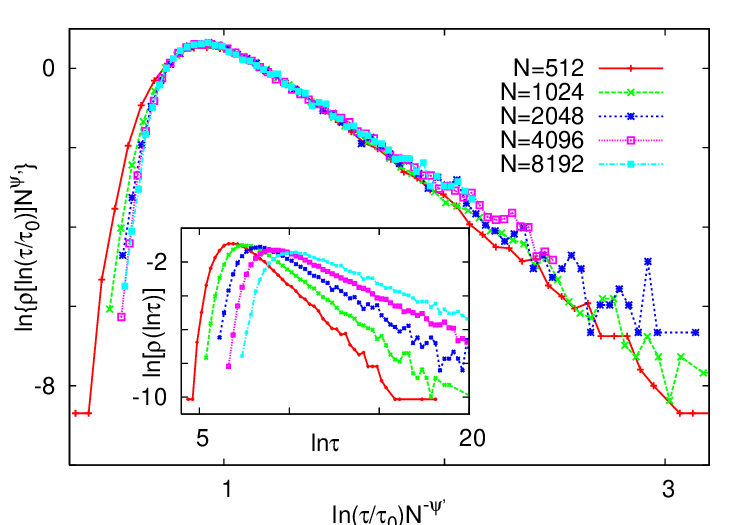}
\caption{
Left: Time-dependence of the spread measured in numerical simulations for $\beta=1$,
at $\lambda=3.410$. The solid line is a fit to the data, yielding the 
parameters $\ln t_0=2.4(4)$ and $\psi=0.53(4)$. 
Right: 
Scaling collapse of the histogram of logarithmic lifetimes obtained for
different sizes $N$. 
For each $N$, the lifetime has been measured in $10^4-10^5$ random samples,
once in each sample. 
Optimal collapse is achieved by the parameters $\ln t_0=3$, $\psi'=0.23$. 
The inset shows the unscaled histograms.
\label{rr1} 
}
\end{figure}

\begin{figure}[h]
\includegraphics[width=0.5\linewidth]{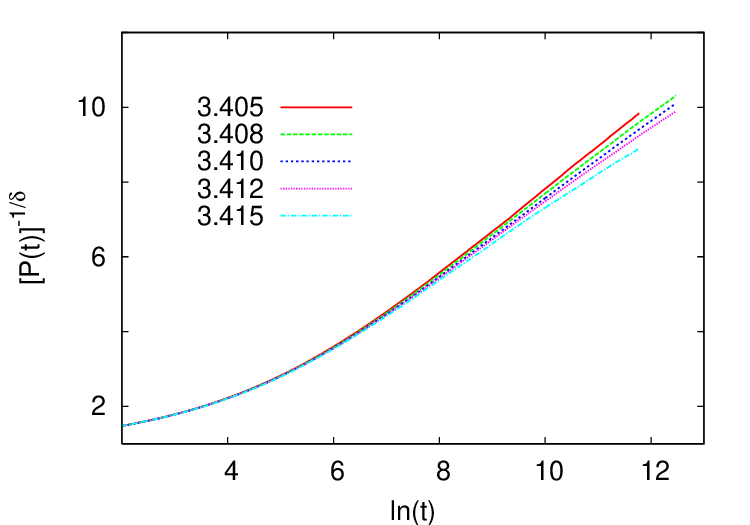}
\includegraphics[width=0.5\linewidth]{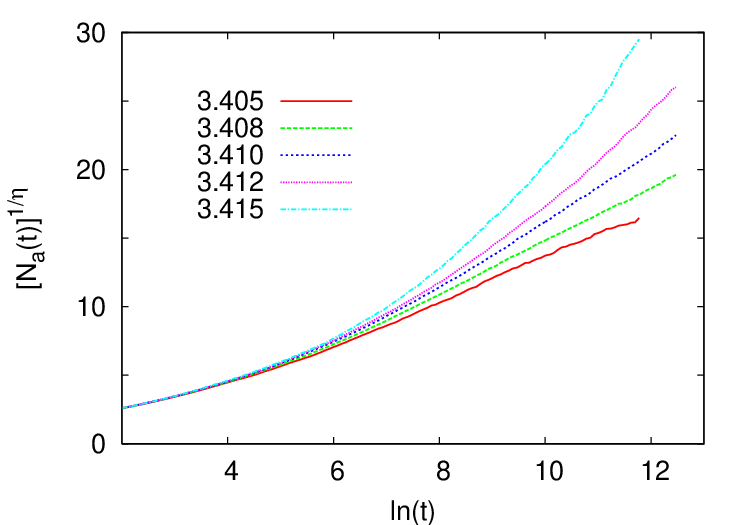}
\caption{
Time-dependence of the survival probability and the average number of active
nodes for $\beta=1$. 
According to Eqs. (\ref{Pt},\ref{Nt}), the asymptote of the critical 
curve must be a line. The estimates of the critical exponents
$\overline{\delta}=1.94$, $\overline{\eta}=0.54$ have been calculated from the data in Table
\ref{table} using Eq. (\ref{deltax}).    
\label{ptnt} 
}
\end{figure}

The tunneling exponent $\psi$ has been determined from the dependence of the
spread on time by fitting a function 
$R(t)=a[\ln(t/t_0)]^{1/\psi}+b$ to the numerical data for times $\ln t>5$. 
For an illustration, see Fig. \ref{rr1}. 
The estimates for various $\beta$ can be found in Table \ref{table}. 
For $\beta=1$, we have determined the tunneling exponent also by measuring the
lifetime $\tau$ starting from a fully active initial state in finite systems. 
Constructing the histogram of $\ln\tau$ for different values of $N$, the exponent
$\psi$ can be determined, according to Eq. (\ref{taudist}), by finding 
an optimal scaling collapse of the data, see Fig. \ref{rr1}. 
The obtained estimate ($\psi'=0.23$) is compatible with that obtained from the time-dependence of the
spread. As can be seen in Fig. \ref{ptnt}, 
the dependence of the survival probability and the average number of active
nodes on time is in agreement with the logarithmic scaling laws given in Eqs. (\ref{Pt},\ref{Nt}). 

In addition to this, we have performed simulations and measurements for model
without parameter disorder (i.e. formally $c=0$), as well, in the same way as
for the model with parameter disorder. The obtained estimates are shown in the last three columns of Table \ref{table}.

In order to make sure about the irrelevance of the normalization of the infection rates by the degree of the source node, we have carried out Monte Carlo simulations of the SIS model on LRC networks. 
The simulations have been realized as follows.  
As in the case of the contact process, i.i.d. random variables  $w_{ij}=w_{ji}$, which have the probability density given in Eq. (\ref{wdist}), 
are assigned to each link. 
Let $N_a(t)$ and $N_l(t)$ denote the number of active nodes and the number of directed links with an active source node (called {\it active links}), respectively, at the beginning of a Monte Carlo step.
Then, with the probability  $N_a(t)/[N_a(t)+\lambda N_l(t)]$, an active node is chosen equiprobably and set to inactive while, 
with the complementary probability 
 $\lambda N_l(t)/[N_a(t)+\lambda N_l(t)]$, an active link is 
chosen equiprobably and its target node is attempted to be activated with 
the probability $w_{ij}$. 
One Monte Carlo step consists of $N_a(t)+\lambda N_l(t)$ 
such updates on average. 
The simulations have been performed with the parameters  $c=0.5$, $w_0=0.2$ on networks with $\beta=1$. 
The critical point is estimated to be at $\lambda_c=0.8877(5)$, 
where we obtained $\overline{\eta}/\overline{\delta}=0.28(2)$. This agrees with the estimate obtained in the contact process for the same $\beta$, confirming the assumption that the distinction between the two models is irrelevant.

\section{Discussion}
\label{discussion}

\begin{figure}[h]
\includegraphics[width=0.5\linewidth]{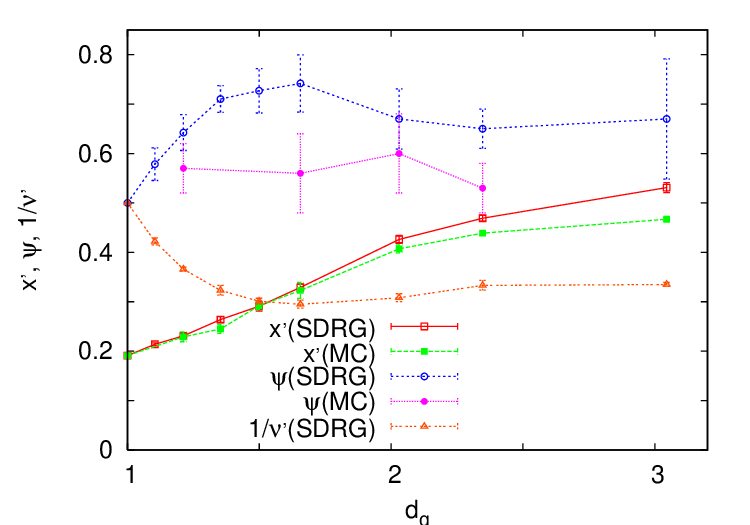}
\includegraphics[width=0.5\linewidth]{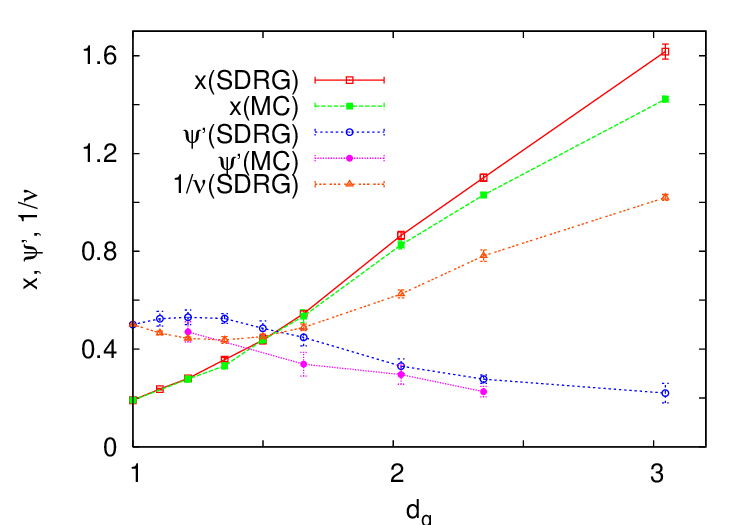}
\caption{
Estimates of different exponents plotted against the graph dimension. 
\label{xdg} 
}
\end{figure}

We have studied the critical contact process and the 
slightly different SIS model in the 
presence of parameter disorder on finite-dimensional random LRC networks 
by an SDRG method and by Monte Carlo simulations.
We have seen by both methods that the behavior of different
observables are compatible with the scaling laws in infinite randomness fixed
points and have obtained estimates of the critical exponents.
As can be seen in Fig. \ref{xdg}, they vary smoothly with the graph
dimension of networks. 
We have observed the infinite-randomness scenario all the way to 
the largest value of
$d_g$ considered in this work and, taking into account the smooth change of
critical exponents with $d_g$ we conjecture this scenario to persist to
arbitrarily large, finite graph dimensions. 

In addition to the model with parameter disorder we performed simulations of
the model with homogeneous parameters. 
Comparing the estimates of the critical exponents obtained in the two cases
(see Table \ref{table}) we can see that they agree within the error of 
measurements. 
This suggests that, at least in the model under study, 
the presence of an additional parameter disorder does not
alter the critical exponents of the infinite randomness
fixed point induced solely by the topological disorder.  
For supporting this finding, a non-rigorous argument based on
the SDRG approach of the problem can be provided, as follows.  

Let us consider the SIS model and the contact process without 
parameter disorder.
All recovery rates are now equal and, in the former, even the infection rates
are uniform.
The SDRG procedure, as formulated for parameter disorder, 
investigates small blocks 
(consisting of two nodes in case of the fusion step) and
tries to replace them by a single effective node. 
The success of the method lies (among others) in that a {\it local control
parameter}, which is determined by the relative magnitude of local recovery and
infection rates within the blocks exists.  
This is much different for topological disorder. 
Here, the locally supercritical regions are expected to be those sub-graphs
which have an over-average number of internal links. 
These regions cannot be detected
simply by investigating and reducing two-node blocks sequentially in the way
the procedure works for parameter disorder. 
Instead larger blocks (at least three nodes) should be investigated 
so that the clusters which are super-critical due to the large number 
of internal links can be found 
\footnote{This problem also arises in the case of the contact process
 on diluted hypercubic lattices. 
Dilution and parameter disorder have been observed to
be in the same universality class \cite{vfm}.}. 
Such a hypothetical SDRG method would be certainly 
much more complicated than the naive
one but the difficulties are merely of technical nature and not 
principal ones.
So we can assume that a renormalization procedure exists by which 
richly linked sub-graphs are selected and replaced by a single node with
a properly calculated effective recovery rate. 
Obviously, the application of this pre-renormalization leads to 
that parameters of the model become disordered. 
In other words, we expect the topological disorder of the
original model inducing parameter disorder in 
the coarse-grained (renormalized) model.
Performing the hypothetical SDRG method up to a finite length scale, 
the resulting
coarse-grained model will therefore be suitable for an SDRG analysis by the
naive method developed for parameter disorder. 
Considering that 
the pre-renormalization carried out up to a finite-scale
does not influence the asymptotical properties of the geometry of the
underlying network and keeping in mind the universality in an IRFP, 
we conclude that 
the additional parameter disorder will not alter the critical behavior with
respect to those induced by the topological disorder.  

Let us now return to the comparison of the SDRG and Monte Carlo
results. 
Concerning the exponent $x'$, the data obtained by the two methods are close
to each other for small $d_g$ but the deviation is significant and
increasing for larger $d_g$, see Fig. \ref{edx}.
However, for larger $d_g$, the effective
strength of disorder is weaker, which manifests itself in that 
$\psi'(d_g)$ is decreasing with $d_g$, and  
this makes the estimates by both methods less
reliable for larger graph dimensions.  
Regarding that the agreement is quite good for moderate $d_g$, we 
conjecture that both models are in the same universality class for any
$d_g$ and attribute the discrepancy for larger $d_g$ to possible 
systematic errors related to the finite size of networks used in the numerical analysis
\footnote{Note that deviations between the SDRG and Monte Carlo estimates have
  been established also in the case of the contact process on hypercubic
  lattices with parameter disorder \cite{vojta3d}.}.
As can be seen in Fig. \ref{xdg}, the exponents $x'$ and $\nu'$ vary with
the graph dimension slowly for large $d_g$ and the exponents $\nu'$ fulfill
the rigorous bound $\nu'\geq2$ for all $d_g$ \cite{chayes}. 
The dependence of these exponents on $d_g$ is similar to that on the 
dimension in hypercubic lattices \cite{ki}. 

The tunneling exponent $\psi$ is found to saturate for large graph dimensions
according to both methods, see Table \ref{tableRG} and \ref{table}, although 
the values obtained by the SDRG method are systematically higher than those 
obtained by Monte Carlo simulations. 
For the exponent $\psi$, one can easily establish an upper bound for 
arbitrary $\beta$, as follows. 
According to  Eq. (\ref{Rt}), 
the exponent $\psi'$ governs the scaling of the typical time $t$ during which 
the activity spreads (in surviving trials) in a finite sample of size $N$ 
from the origin to the node in the Euclidean distance $N/2$, through   
$\ln t\sim N^{\psi'}$. 
Furthermore, increasing $\beta$ amounts to that the number of 
long links increases in the network. 
Then, it is plausible to assume that the time $t$ must not increase with
increasing $\beta$ since the long edges promote the spreading of activity. 
Consequently, the exponent $\psi'(\beta)$ must not increase with $\beta$. 
Since $\psi'(\beta=0)=1/2$, it follows that 
\be 
\psi'(\beta)\le 1/2
\ee
for any $\beta$.

As can be seen from the data, the decreasing tendency of $\psi'(\beta)$ and
the above inequality are fulfilled except of the
SDRG estimates for small values of $\beta$; nevertheless, the upper bound lies 
within the error of estimates also here. 
This may be attributed to the relatively strong finite-size
corrections, depending heavily on the chosen distribution of the parameters. 
In order to have more accurate results, even larger system sizes should be 
needed, or, possibly, other forms of disorder, where the finite-size corrections have a different sign. 

It is interesting to note that the possible saturation of
$\psi$ with increasing dimension has been observed also on hypercubic
lattices, although to a value ($0.46$) lower than that obtained 
by either methods of this work on LRC networks \cite{ki}.    
    
The exponents describing the critical dynamics of the contact process, 
$\overline{\delta}$ and $\overline{\eta}$, can be calculated from the data in
Table \ref{tableRG} and \ref{table}. Both vary with the graph dimension, 
the former increasing, the latter decreasing with $d_g$, in agreement with the
tendencies on hypercubic lattices.  

\section{Conclusions and outlook}
\label{conclusions}  

We have studied in this work the contact process, as well as the SIS model on
LRC networks, where the topological disorder leads to infinite-randomness critical behavior. 
We have found by a numerical analysis supported by heuristic argumentations, 
that an additional parameter disorder does not change the critical exponents
of the transition. In this way, the exponents are universal in the sense that
they are determined exclusively by the topology of the network. 
This result opens up an alternative way of investigating infinite randomness
critical behavior induced by topological disorder in general. 
Namely, after introducing parameter disorder, which is irrelevant in the above
sense, one can apply the efficient SDRG method to the model.  

Furthermore, the present numerical study has shown that in the case it is
questionable whether the phase transition in a given model is of IRFP type or
conventional one, it is better to concentrate on the critical behavior rather
than searching for a Griffiths phase (an accompanion of IRFP), which may be,
in case of a relatively weak disorder (small $\psi'$), very hard to
detect.      

We have clearly seen in the present work, that the critical exponents
smoothly vary with the topological dimension and, 
for moderate values of $d_g$ we provided estimates on them. 
We have obtained from these data indications on the possible limiting behavior
of these exponents when $d_g\to\infty$. 
Although the exponents for finite $d_g$ does not seem to be determined 
exclusively by $d_g$ (cf. the estimates on hypercubic lattices \cite{ki}), 
it is an intriguing question whether the limiting values of them 
are universal for $d_g\to\infty$.
In a wider context, 
it is also a challenging question, whether there are such 
topological characteristics of the underlying networks 
which determine the critical exponent of the IRFP on the top of them
unambiguously or at least approximately.  

As mentioned in the Introduction, the transverse-field Ising model is in the
same universality class as the contact process, in the presence of parameter
disorder. 
Our results indicate that the critical behavior of the 
above model defined on LRC
networks where there is exclusively topological disorder would be also described by 
an IRFP characterized by critical exponents obtained in the present work.

\ack
This work was supported by the J\'anos Bolyai Research Scholarship of the
Hungarian Academy of Sciences, by the National Research Fund
under grant no. K75324, and partially supported by the European Union and the
European Social Fund through project FuturICT.hu (grant no.:
TAMOP-4.2.2.C-11/1/KONV-2012-0013).


\end{document}